\documentclass[12pt]{article}

\usepackage[dvips]{epsfig}

\begin{document}

\title{Scattering of periodic solitons}

\author{$\Re$J Cova\footnote{rcova@luz.ve}  \\ Dept
de F\'{\i}sica, FEC \\ La Universidad del Zulia \\ Apartado 15332 \\
Maracaibo 4005-A \\ Venezuela \\
\and WJ Zakrzewski\footnote{W.J.Zakrzewski@durham.ac.uk} \\ 
Dept of Mathematics \\ University of Durham \\
Durham DH1 3LE \\ UK}

\maketitle


\abstract{With the help of numerical simulations we study
$N$-soliton scattering ($N$=3,4) in the (2+1)-dimensional $CP^1$
model with periodic boundary conditions. When the solitons are scattered from
symmetrical configurations the scattering angles observed agree with
the earlier $\pi/N$ predictions based on the model on
$\Re_2$ with standard boundary conditions. When the initial configurations are not 
symmetric the angles are different from $\pi/N$. We present an explanation of our
observed patterns based on a properly formulated geodesic approximation.}

\section{Introduction \label{sec:intro}}

Physics in (2+1) dimensions is an area of much active research, covering
topics that include Heisenberg ferromagnets, the quantum Hall effect,
superconductivity, nematic crystals, topological fluids, vortices and
solitary waves \cite{cho}.  Most of these systems are non-linear. In their
mathematical description the well-known family of sigma models plays a
starring role. The simplest Lorentz-covariant soliton model in (2+1)
dimensions is the $CP^1$ or non-linear $O(3)$ sigma model. Its solutions,
sometimes called `lumps', are
realisations of harmonic maps, a long-established area of research in pure
mathematics. However, analytical $O(3)$ solitons solutions have only been found for
the static case; the full time-dependent model must be studied
using numerical methods and/or other approximation procedures \cite{leese}.

Sigma models are also useful as low dimensional analogues of field theories
in higher dimensions. In effect, the $O(3)$ model in two dimensions
exhibits conformal invariance, spontaneous symmetry breaking, asymptotic
freedom and topological solitons, properties similar to those present in a
number of important field theories in (3+1) dimensions --like
the Skyrme model of nuclear physics \cite{skyrme}. 

We are concerned with the planar $CP^1$ model (both in its original and 
Skyrme-like versions) with periodic boundary conditions where the solitons are 
harmonic maps
\( T_2 \mapsto S_2 \). A rich diversity of phenomena has been found in this
model \cite{nonl,martin,epj2k1}, going beyond the two-soliton and annular
structures one might expect by analogy with the model with standard boundary 
conditions, where the soliton fields are harmonic maps 
\( \Re_{2} \cup \{\infty\} \approx S_2 \mapsto S_2 \).

For planar systems, $N$ identical lumps initially placed 
at the vertices of a rectangular $N$-gon, fired with equal speed to collide head-on
at the centre of the polygon, scatter and emerge on the vertices of the dual
polygon. Such $\pi/N$ scattering, studied for the $CP^1$ model on $\Re_2$ 
in reference \cite{pin}, may be understood on
symmetry grounds: the initial data has $D_N$ (dihedral group) symmetry and the 
time evolution respects that. But imposing periodic 
boundary conditions breaks the foresaid symmetry, and the interesting question
whether dual-polygon scattering still holds for this case must be investigated.

Through numerical simulations
on $T_2$ we confirm the $\pi/N$ scattering for 3 and 4 
identical lumps. The case $N=2$ has being considered elsewhere \cite{nonl} 
and it conforms to the well-documented scattering at 90$^{\circ}$.
We also look into non-symmetrical configurations (they do not scatter at $\pi/N$)
and explain the results using the geodesic approximation. 

In the following section we introduce our periodic $CP^1$
model. The numerical procedure is explained in section
\ref{sec:numerical}. Section \ref{sec:3} analyses collisions between three
solitons, and the case  $N=4$ is considered in section \ref{sec:4}.
In section \ref{sec:cc} we present our version of the geodesic
approximation which includes the case of initial 
configurations that are not symmetrical. We find that our
predictions (based on this approximation) are in agreement with what is
observed in numerical simulations. We close with some concluding
remarks in section \ref{sec:conclusions}.

\section{The $CP^1$ model on the torus \label{sec:model}}

The non-linear $O(3)$ model involves three real scalar fields
\( \vec{\phi}(x^{\mu})\equiv (\phi_{1},\phi_{2},\phi_{3})\, \)
which satisfy the constraint that $\forall$
\( x^{\mu} \equiv (x^{0},x^{1},x^{2})=(t,x,y) \in T_2 \)
the fields lie on the unit sphere $S_{2}$:
%
\begin{equation}
\vec{\phi}.\vec{\phi}=1.
\label{eq:constraint}
\end{equation}
%
Subject to this constrain,  the Lagrangian density of the system is
%
\begin{equation}
{\cal L}=\frac{1}{4}
(\partial_{\mu}\vec{\phi}).(\partial^{\mu}\vec{\phi}).
  \label{eq:purelagphi}
\end{equation}    
The invariance of the model under global $O(3)$ rotations in field-space is apparent.
%
%
%

The model is conveniently recast in terms of one independent
complex field $W$ (the $CP^1$ formulation) related to $\vec{\phi}$ 
via
%
\begin{equation}
\vec{\phi}=(\underbrace{\frac{W+\bar{W}}{|W|^{2}+1}}_{\phi_1},
\underbrace{i\frac{-W+\bar{W}}{|W|^{2}+1}}_{\phi_2},
\underbrace{\frac{|W|^{2}-1}{|W|^{2}+1}}_{\phi_3}).
  \label{eq:projection}
\end{equation}     
The Lagrangian (\ref{eq:purelagphi}) now reads 
%
\begin{equation}
{\cal L}=
\frac{|\partial_{t}W|^{2}-2|\partial_{z}W|^{2}-
2|\partial_{\bar{z}}W|^{2}}{(1+|W|^{2})^{2}}, 
\label{eq:purelagw}
\end{equation}                               
%
%
%
where $z=x+iy$ and $\bar{z}=x-iy$. 
For all $t$, the fields $W$ are mappings from the torus $T_{2}$ to
the sphere $S_{2}$, {\em ie,} they satisfy the periodic boundary conditions
%
\begin{equation} 
W(z+mL+inL)=W(z),
  \label{eq:boundaryw}
\end{equation}
where $m,n=0,1,2,\ldots$ and the period $L$ denotes the size of
the square torus.  

The \emph{static} soliton solutions of the model 
are thus doubly periodic functions of $z$, that is, elliptic
functions that may be expressed through Weierstrass $\sigma$-function as
\cite{goursat,lawden}
%
\begin{equation}
W= \lambda \prod_{j=1}^{\kappa}
\frac{\sigma(z-a_{j})}{\sigma(z-b_{j})}, \qquad 
\label{eq:instanton}
%
%
\sum_{j=1}^{\kappa} a_{j}=\sum_{j=1}^{\kappa} b_{j}.
\end{equation}         
The complex number $\lambda$ is related to
the overall size of the solitons, the zeros $a_j$ and poles $b_j$
determine their sizes and positions on $T_2$, and the positive
integer $\kappa$ is the order of the elliptic function $W$.

%
%

%
%
%

\begin{figure} 
\mbox{\epsfig{file=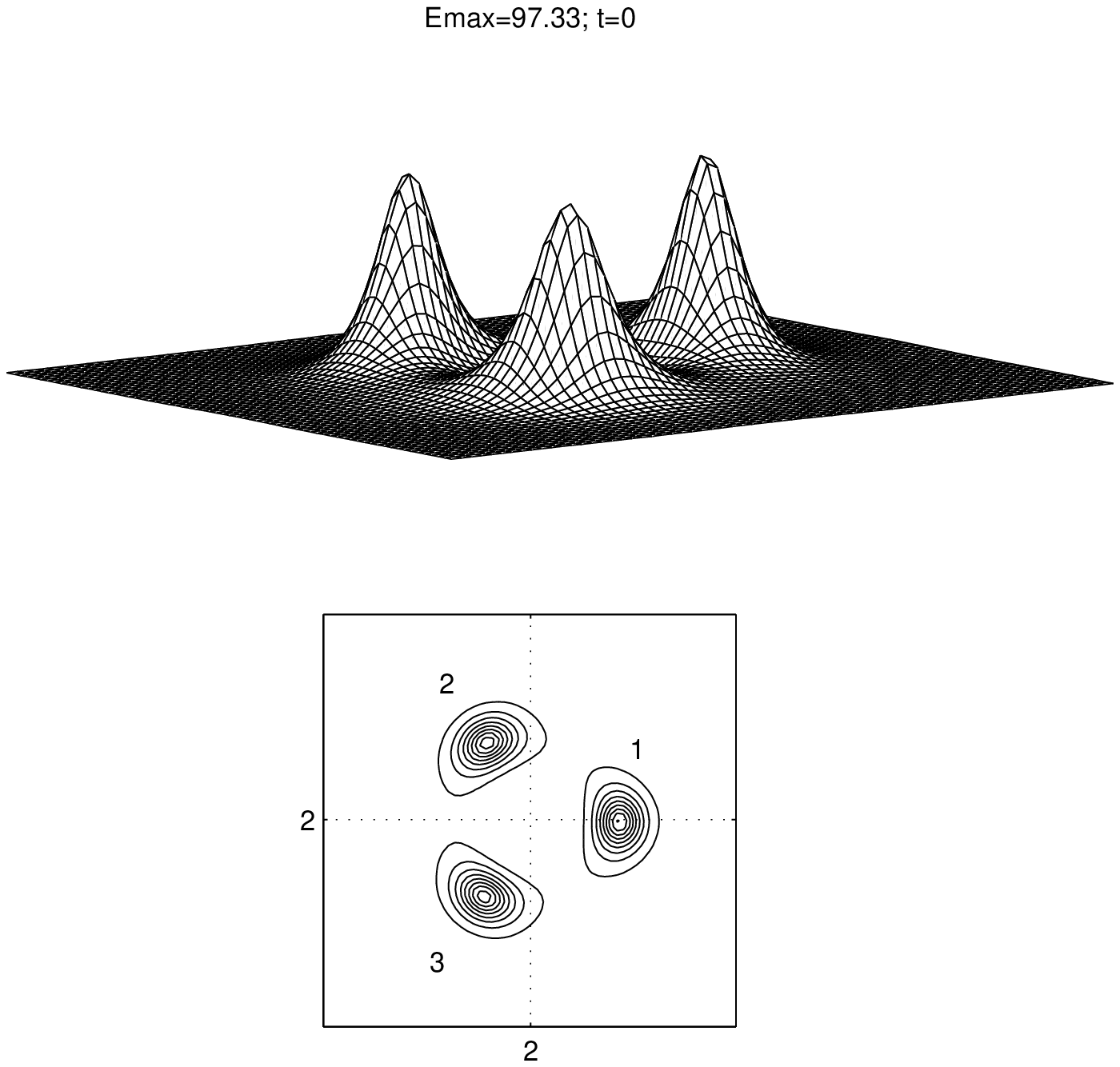}} 
\caption{The energy distribution for $N$=3 at the initial 
time, both in three dimensional and contour plot forms.}
\label{fig:pic1n}
\end{figure}

The static energy density (or potential energy density) 
associated with (\ref{eq:instanton}) can be read-off from 
(\ref{eq:purelagw}):
%
%
\begin{equation}
E=
2\frac{|\partial_{z}W|^{2}+|\partial_{\bar{z}}W|^{2}}
{(1+|W|^{2})^{2}}.
\label{eq:energy}
\end{equation}
Pictures of this energy distribution reveal the familiar $CP^1$ lumps,
as those of figure \ref{fig:pic1n} for example. 
 
The energy is related to the topological charge by the Bogomolnyi bound
\begin{equation}
E \geq 2\pi |N|.
\label{eq:bound}
\end{equation}    
The instanton solutions correspond to the equality in
(\ref{eq:bound}):
solutions carrying $N>0$ ($N<0$) imply $\partial_{\bar{z}}W=0$
($\partial_{z}W=0$), the Cauchy-Riemann conditions for $W$ being an
analytic function of $z$ ($\bar{z}$). Note that the simplest non-trivial 
elliptic function is of order two, hence there are no single-soliton solutions on
$T_2$.

We utilise the pure model (\ref{eq:purelagw}) for $N=3$, but for $N=4$ the 
energy involved is so large that the well-known instability of the planar model
(recall its conformal invariance means that the lumps can have any width)
breaks the numerical procedure fast. A stabilising Skyrme term must be introduced 
for this case. 

%

\section{Numerical procedure \label{sec:numerical}}

In this paper we want to discuss time dependent solutions of
the model (\ref{eq:purelagw}) and its Skyrme version [see equation 
(\ref{eq:skylagw})], where
the time dependence describes the movement of solitons. As we do not have
analytical time-dependent solutions, we resort to numerical 
simulations. For this we take 
fields of the form (\ref{eq:instanton}) as the initial conditions.
Since during the simulations the field $W$ may become
arbitrarily large, we have preferred to run our simulations in the
$\phi$-formulation of the model.
%
%
One returns to the $CP^1$ formulation inverting (\ref{eq:projection}):
\linebreak
%
$W=(1-\phi_{3})/(\phi_{1}+i\phi_{2})$.

Strictly speaking, truly independent solitons can only be obtained in the
asymptotic regime of large soliton separation, which really never happens
on a compact manifold such as $T_2$. However, each factor
$W_j\equiv\sigma(z-a_j)/\sigma(z-b_j)$ in (\ref{eq:instanton}) (when
$a_j\sim b_j$) roughly represents one soliton, providing a setting to
studying more or less independent structures.  The present work is limited
to systems in the topological classes 3 and 4. These systems move, collide
and scatter off upon being set into motion by boosting each $W_j$
separately.  The initial-value problem is then completely specified by
giving both $W(t)$ and $\partial_t W(t)$ at the initial time $t=0$.

For a square torus we have the so-called lemniscatic
case \cite{mcgraw} where $\sigma$ posseses the simple
Laurent expansion 
%
\begin{equation}
\sigma(u)=\sum_{j=0}^{\infty}{G}_{j}u^{4j+1},
\qquad G_j \equiv G_j(L) \in \Re.
\label{eq:sig}
\end{equation}        
For the accuracy of our calculations it is sufficient to compute 
the series (\ref{eq:sig}) up to $G_5$ (our coefficients for $j>5$
are negligible):
%
\begin{displaymath}
\left.
\begin{array}{lll}
G_{0}&=&1 \\
G_{1}&=&-0.7878030 \\
G_{2}&=&-0.221654845 \\
G_{3}&=&9.36193 \times 10^{-3} \\
G_{4}&=&7.20830 \times 10^{-5} \\
G_{5}&=&2.37710 \times 10^{-5}
\end{array}
\right\}.
\end{displaymath}
 We have employed the fourth-order Runge-Kutta method and
approximated the spatial derivatives by finite differences.  The
Laplacian has been evaluated using the standard nine-point formula
and, to further check our results, a 13-point recipe has also been
used. Respectively, the Laplacians are:
%
%
%
%
\begin{equation}
\nabla^2 = \frac{\left[
 \begin{array}{cccc} 
  1 & 4   & 1  \\
  4 & -20 & 4 \\
  1 & 4   & 1
\end{array}
\right]}
{6 \times (\delta x)^2}, \qquad
%
%
\nabla^2= \frac{\left[
 \begin{array}{ccccc}
    &    & -1  &    &   \\
    & 1 & 12 & 1 &    \\
  -1 & 12 & -48  & 12 & -1  \\
    & 1 & 12 & 1 &   \\
    &    & -1  &    &
\end{array}
\right]}
{10 \times (\delta x)^2}. 
\end{equation}

The discrete model has been evolved on a \( n_x \times n_y = 200
\times 200 \) square periodic lattice with spatial and time steps $\delta
x$=$\delta y$=0.02 and $\delta t$=0.005, respectively. The vertices
of the \underline{fundamental period cell} we have used for our
simulations were at
%
\begin{equation}
(0,0),\; (0,L),\; (L,L),\; (L,0), \quad L=n_x \times \delta x=4.
\label{eq:cell}
\end{equation}

Unavoidable round-off errors have gradually shifted the fields away
from the constraint \( \vec{\phi}.\vec{\phi}=1 \). So, like in the
planar case \cite{chaos}, to correct for this
 we have rescaled 
$
\vec{\phi} \rightarrow \vec{\phi}/\sqrt{\vec{\phi}.\vec{\phi}}
$
every few iterations.  Each time, just before the rescaling
operation, we have evaluated the quantity \( \mu \equiv
\vec{\phi}.\vec{\phi} - 1 \)\, at each lattice point. Treating the
maximum of the absolute value of $\mu$ as a measure of the numerical
errors, we have found that max$|\mu|$ $\approx$ 10$^{-8}$.
 This magnitude is useful as a guide to determine how reliable a
given numerical outcome is. Usage of unsound numerics
in the Runge-Kutta evolution shows itself as a rapid growth of
max$|\mu|$; this also occurs, for instance, when the solitons 
pinch-off.

\bigskip
The parameter $\lambda$ in (\ref{eq:instanton})
has been set to $\lambda=(1,0)$ all throughout.

\section{Three solitons \label{sec:3}}

First we have considered states with three solitons. Our initial
configuration is given by taking $\kappa=3$ in (\ref{eq:instanton}), 
the elliptic function of order 3
%
\begin{equation}
W= \frac{\sigma(z-a_{1})}{\sigma(z-b_{1})}
\frac{\sigma(z-a_2)}{\sigma(z-b_2)} 
\frac{\sigma(z-a_3)}{\sigma(z-b_3)},
\quad \sum_{j=1}^3 a_{j}=
\sum_{j=1}^3 b_{j}.
\label{eq:3instanton}
\end{equation}      
The values of $a_j, b_j$ have been selected in such a way that the
solitons lie symmetrically around a circle in the network
(\ref{eq:cell}). This is easily achieved by fixing $a_1,b_1$ 
to reasonable values and setting 
%
\begin{equation}
\left.
\begin{array}{lll}
a'_1 &=& a_1-c \\
b'_1 &=& b_1-c
\end{array}
\right\},  
\label{eq:primed1}
\end{equation}
where $c=(2,2)$ stands for the centre of the period cell.
Then 
%
\begin{equation}
\left.
\begin{array}{lll}
a'_j &=& a'_1 \exp{(i \beta_j)} \\
b'_j &=& b'_1 \exp{(i \beta_j)}   
\end{array}
\right\}; \quad \beta_j=(j-1)\frac{2\pi}{3}, \; j=1,2,3. 
\label{eq:rotation}
\end{equation}    
This symmetrical arrangement gives solitons of the same size, and
satisfies the selection rule in (\ref{eq:3instanton}) for
any choice of the complex numbers $a'_1, \, b'_1$:
\begin{eqnarray}
\sum_{\kappa=1}^3 a'_j &=& 
                       (0,0) \nonumber \\
                       &=& \sum_{\kappa=1}^3 b'_j.
\label{eq:vaca}
\end{eqnarray}

Next we go back to $a_j$ and $b_j$ through
%
\begin{equation}
\left.
\begin{array}{lll}
a_j &=& a'_j+c \\
b_j &=& b'_j+c
\end{array}
\right.
\label{eq:primedj}
\end{equation}
and supply the system with an initial speed $v_0$ by boosting
%
\begin{equation}
\left.
\begin{array}{lll}
a_j & \rightarrow & a_j + v_j t \\
b_j & \rightarrow & b_j + v_j t
\end{array}
\right\}; \quad v_j=-v_0 \exp{(i \beta_j)}.     
\label{eq:boost}
\end{equation}
It is now possible to evaluate the time derivative of $W$ at the 
initial time. Inserting (\ref{eq:boost}) into (\ref{eq:3instanton})
we get 
%
\begin{equation}
\partial_t W(t)\mid_{t=0} = - \prod_{j=1}^{3} v_j
[\frac{\partial_z \sigma(z-a_j)}{\sigma(z-a_j)}-
\frac{\partial_z \sigma(z-b_j)}{\sigma(z-b_j)}] W_j,
\label{eq:wt}
\end{equation}
where $W_j \equiv \sigma(z-a_j)/\sigma(z-b_j)$
denotes the $j$-th soliton. Our initial-value problem
is defined by (\ref{eq:3instanton}) and (\ref{eq:wt}).

\bigskip
Choosing  
%
%
\begin{equation}
\left.
\begin{array}{lll}
a_1 = (3, 2) \\
b_1 = (1.45, 1.95) \\  
v_0 = 0.35
\end{array}
\right\},
\label{eq:3values}
\end{equation}     
gives an initial configuration whose energy density 
is exhibited in figure \ref{fig:pic1n}. The solitons are placed on the
vertices of an equilateral triangle; the first lump is situated 
along the line $y$=2 at $\beta_1=0^{\circ}$, with 
respect to which solitons 2 and 3 are rotated $\beta_2=120^{\circ}$ and 
$\beta_3=240^{\circ}$, respectively. These angles are readily checked 
from the picture with the help of a protractor. 

\begin{figure} 
\mbox{\epsfig{file=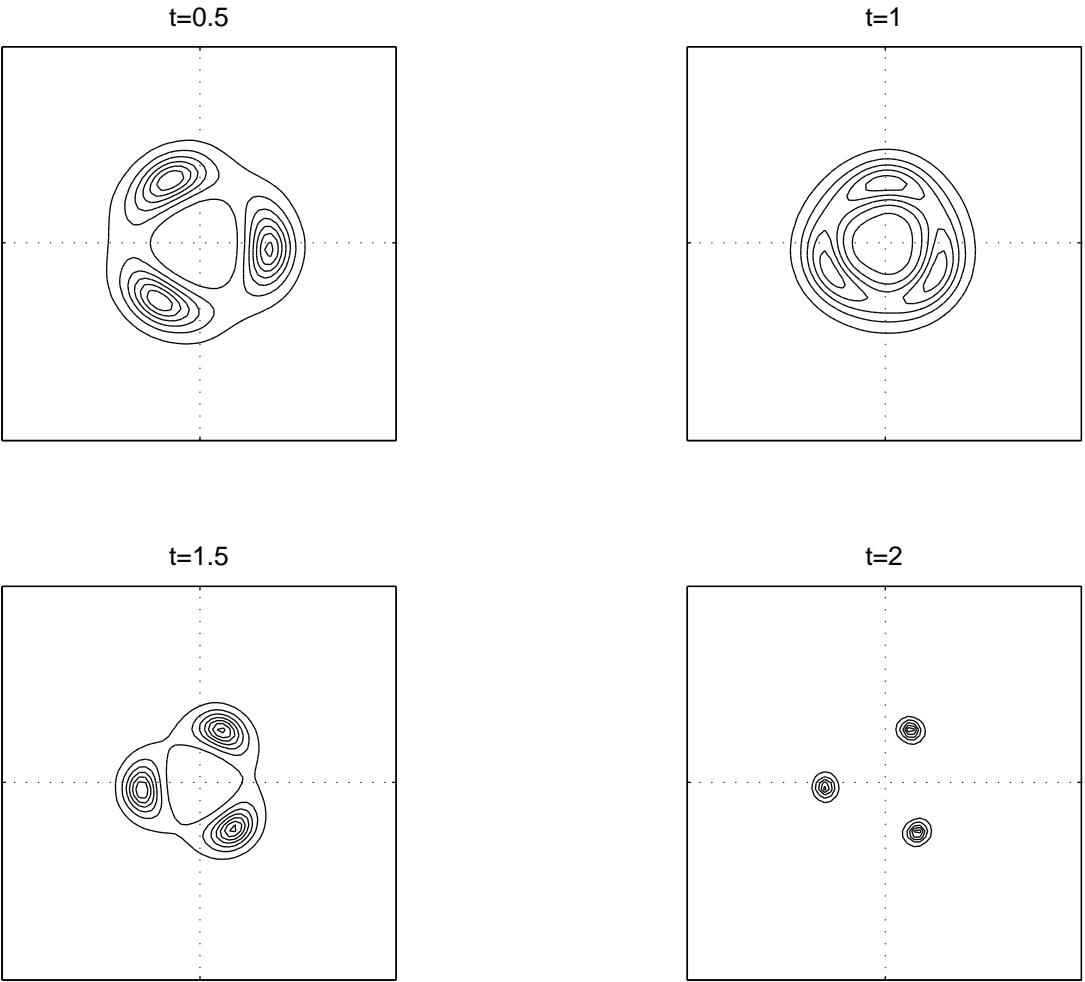}}
\caption{The evolution of the three lumps of figure 
\ref{fig:pic1n} at various times.}
\label{fig:pic2n}
\end{figure}

\begin{figure}
\mbox{\epsfig{file=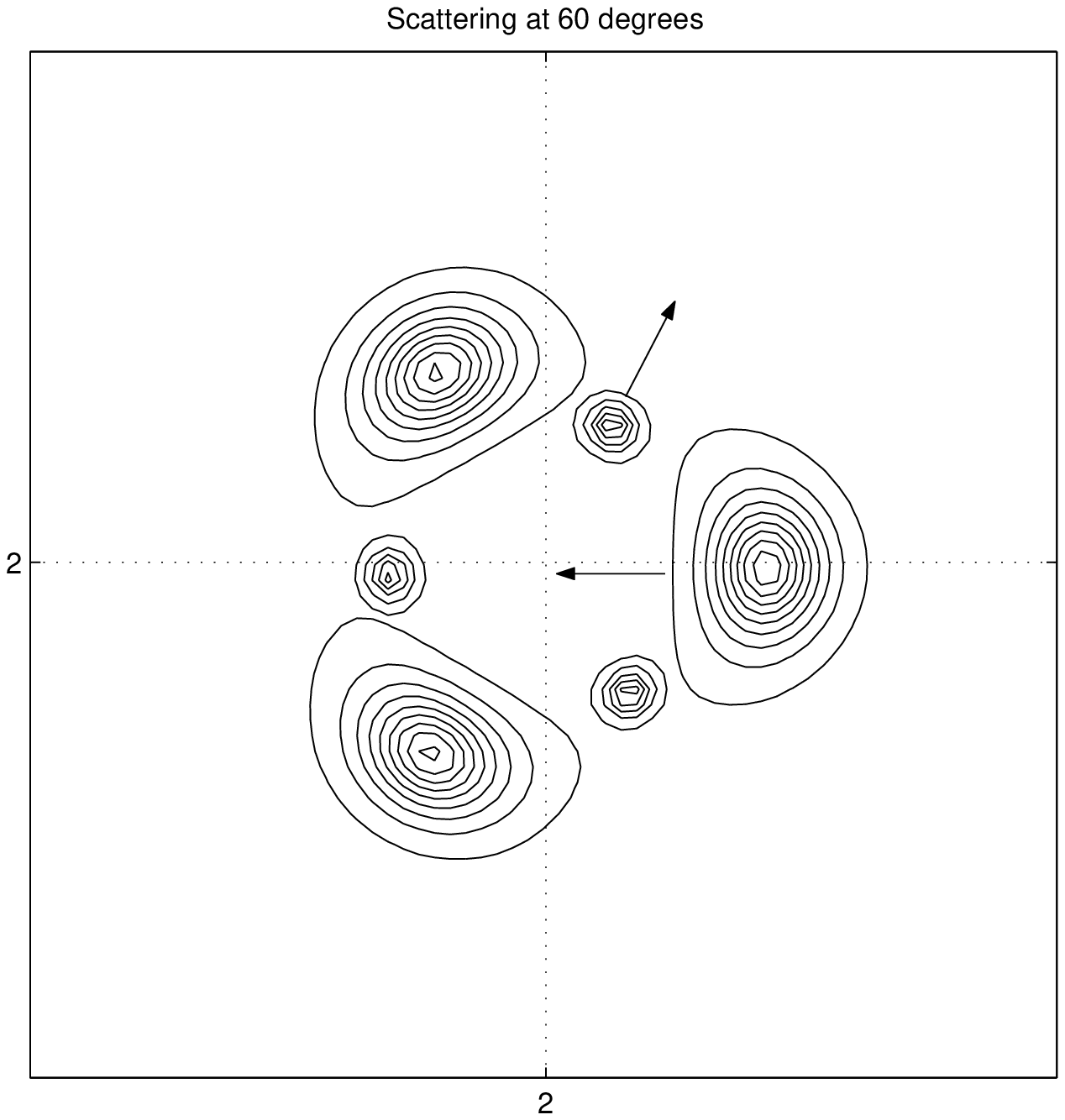}}
\caption{Scattering at $\pi/3$ radians.}
\label{fig:pic3n}
\end{figure}

The results of simulations for this case are depicted in figure
\ref{fig:pic2n} for various times. The three solitons are fired
towards the centre of the triangle and in so doing they expand: the peak
$E_{max}$ of the energy density decreases. The solitonic trio
collides head-on and coalesces in a ringish structure, then emerging towards the 
vertices of the dual triangle, that is, the initial line of
approach of a given incoming soliton forms an angle of $\pi/3$ with the
line along which an outgoing, emerging soliton progresses. 
This $\pi/3$ scattering can be best appreciated in figure \ref{fig:pic3n}, 
where both the initial state ($t$=0) and the final state 
($t=2$) are displayed together.

Returning to figure \ref{fig:pic2n} we see that $E_{max}$
becomes narrower with time, particularly after the lumps scatter off and
start drawing away from each other. At $t=2$, for instance, the maximum 
value of the energy density goes up to $E_{max}=579.37$. Soon after $t>2$
this peak gets so spiky that the numerics breaks down: the
instability of the planar $O(3)$ model takes over and leads to
singularity formation.

Our result is noteworthy: the initial configuration, although positioned at the
vertices of an equilateral triangle in the period cell, does not produce $D_3$
symmetry because the torus itself, being homogeneous but not isotropic, has no such
symmetry (the fundamental grid has directed sides).  One could reasonably expect
the lumps to scatter along directions that need not respect $D_3$ symmetry.
Therefore, the rationale applied to explain $\pi/N$ scattering for the model on
$\Re_2$ is no longer valid for the model with periodic boundary conditions.

However, we observe that as the solitons are well localised the boundary
conditions may not be very important. Note that our numerical results
are consistent with this expectation. To test this further we could place 
our solitons nearer the edges of the grid and see whether we still
observe the 60$\sp{0}$ scattering. We hope to investigate this issue 
in the future.

\section{Four solitons \label{sec:4}}

Next we have looked at the $N=4$ configurations. The initial field
is given by the elliptic function of order 4
%
\begin{equation}
W = \frac{\sigma(z-a_{1})}{\sigma(z-b_{1})}
\frac{\sigma(z-a_2)}{\sigma(z-b_2)}
\frac{\sigma(z-a_3)}{\sigma(z-b_3)}
\frac{\sigma(z-a_4)}{\sigma(z-b_4)},
\quad \sum_{j=1}^4 a_{j}=
\sum_{j=1}^4 b_{j},
\label{eq:4instanton}
\end{equation}
where $a_j, b_j$ are chosen so that the solitons sit 
symmetrically at the vertices of a square in the basic cell. A
treatment parallel to that of the previous section leads to
%
\begin{equation}
\left.
\begin{array}{lll}
a'_j &=& a'_1 \exp{(i \varphi_j)} \\
b'_j &=& b'_1 \exp{(i \varphi_j)}
\end{array}
\right\}; \quad \varphi_j=(j-1)\frac{\pi}{2}, \; j=1,2,3,4,
\label{eq:4rotation}
\end{equation}
The condition between the zeros and poles is again verified $\forall 
\; a'_1, \, b'_1$ :
%
\begin{eqnarray*}
\quad \sum_{j=1}^4 a'_{j} 
                          &=& (0,0) \\ 
                          &=& \sum_{j=1}^4 b'_{j}.
\label{eq:rule4}
\end{eqnarray*}    

Both the initial velocity and the time dependence are 
introduced by boosting:
%
\begin{equation}
\left.
\begin{array}{lll}
a_j & \rightarrow & a_j + v_j t \\
b_j & \rightarrow & b_j + v_j t
\end{array}
\right\}; \quad v_j=-v_0 \exp{(i \varphi_j)},
\label{eq:4boost}
\end{equation}
where equation (\ref{eq:primedj}) should be kept in mind.

\bigskip
As pointed out in section \ref{sec:model}, four lumps involve a large energy and
while evolving in time they become too spiky before we can learn anything much
about the
scattering process. We have
therefore studied this system in the Skyrme version of the theory,
where the solitons are stable and may be examined for as long as
required.  

Instead of the Lagrangian density (\ref{eq:purelagw}) we have thus taken:

\begin{equation}
{\cal L}=
\frac{|W_{t}|^{2}-2 |W_{z}|^{2}-2 |W_{\bar{z}}|^{2}}
{(1+|W|^{2})^{2}} 
\nonumber \\
  -8 \theta_{1}\frac{|W_{z}|^{2}-|W_{\bar{z}}|^{2}}
       {( 1+|W|^{2} )^{4}}
     ( |W_{t}|^{2}+|W_{z}|^{2}-|W_{\bar{z}}|^{2} ).
\label{eq:skylagw}
\end{equation}
The configuration (\ref{eq:instanton}) is no longer an
exact solution of the field equation derived from 
(\ref{eq:skylagw}), albeit it is a very good approximation 
to it. We should also stress that the presence of a Skyrme term does not  
affect the trajectory of the lumps before the collision or their scattering angle. 

\begin{figure}
\mbox{\epsfig{file=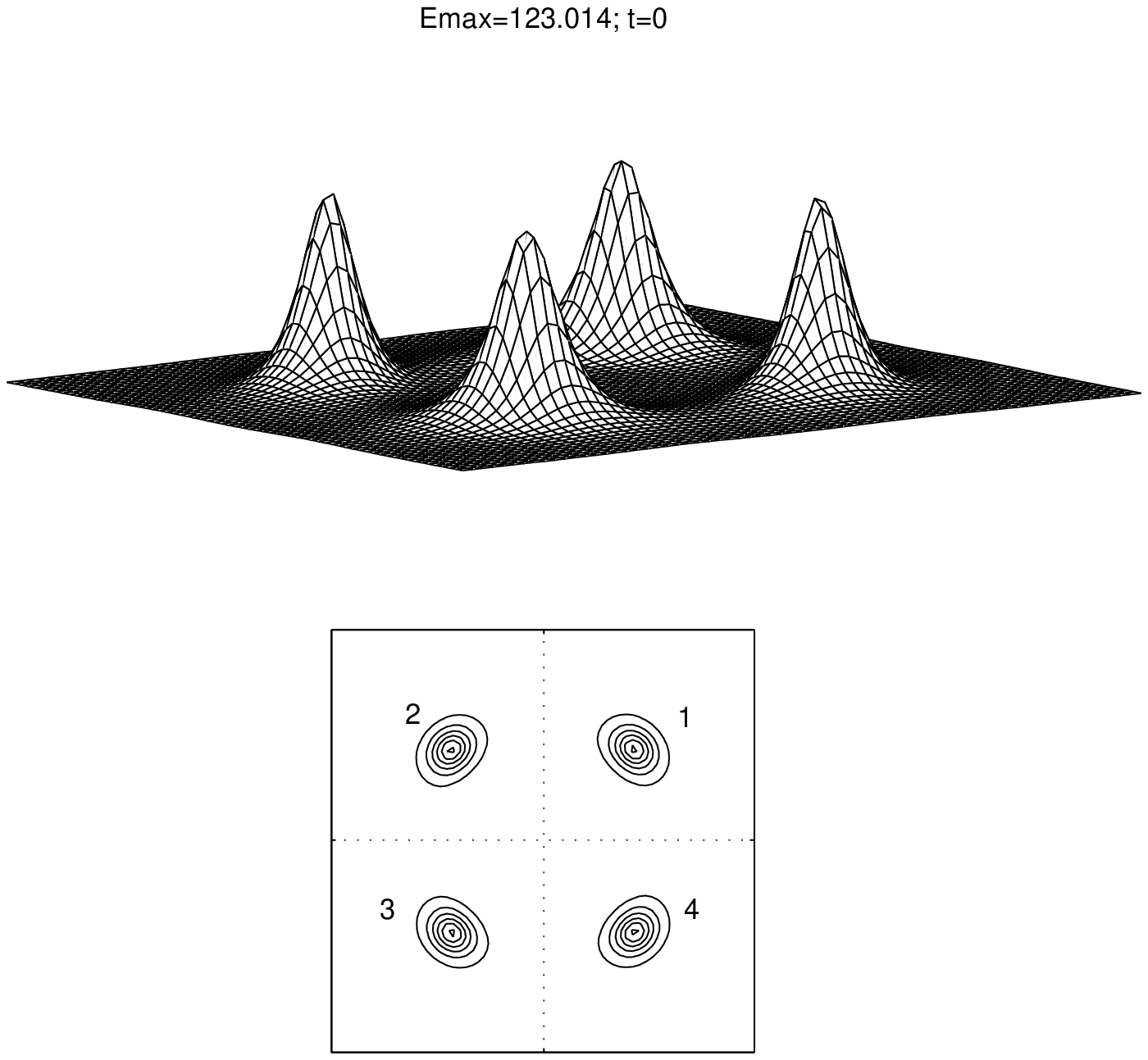}}
\caption{The energy distribution for $N$=4 at $t=0$.}
\label{fig:pic4n}
\end{figure}

\bigskip
Taking
%
\begin{equation}
\left.
\begin{array}{lll}
a_1 = (2.05, 2.55) \\
b_1 = (2.95, 2.95) \\
\theta_1 = 10^{-3} \\
v_0 = 0.35 \exp(i\pi/4) 
\end{array}
\right\},
\label{eq:4values}
\end{equation} 
entails the state of identical `baby skyrmions' illustrated in figure 
\ref{fig:pic4n}. The solitons $W_2$, $W_3$ and $W_4$ are rotated $90^{\circ},
180^{\circ}$ and $270^{\circ}$ with respect to $W_1$, which we have
conveniently placed in the first quadrant, roughly on the central
diagonal joining the grid points (0,0)-(4,4).  

\begin{figure}
\mbox{\epsfig{file=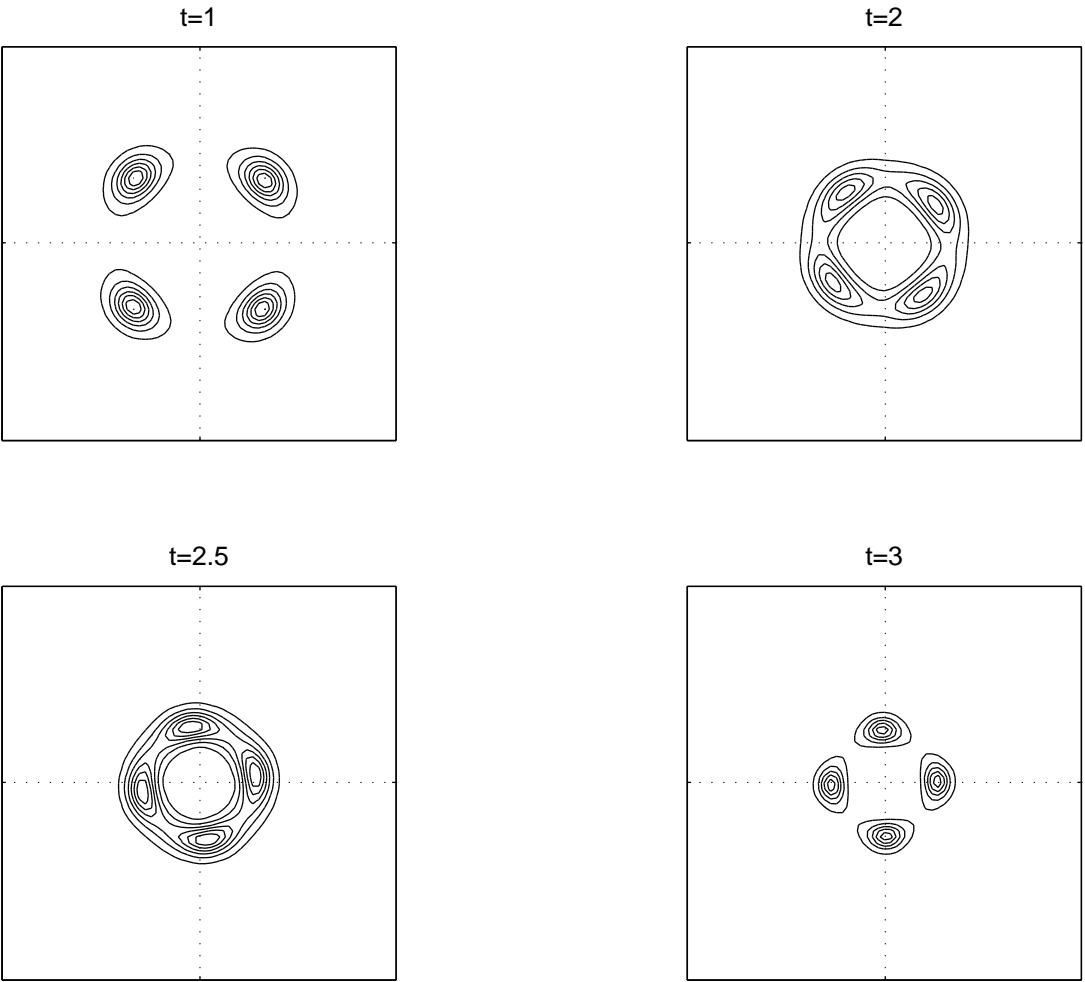}}
\caption{The evolution of the four lumps of figure
\ref{fig:pic4n} gets underway.}
\label{fig:pic5n}
\end{figure}

\begin{figure}
\mbox{\epsfig{file=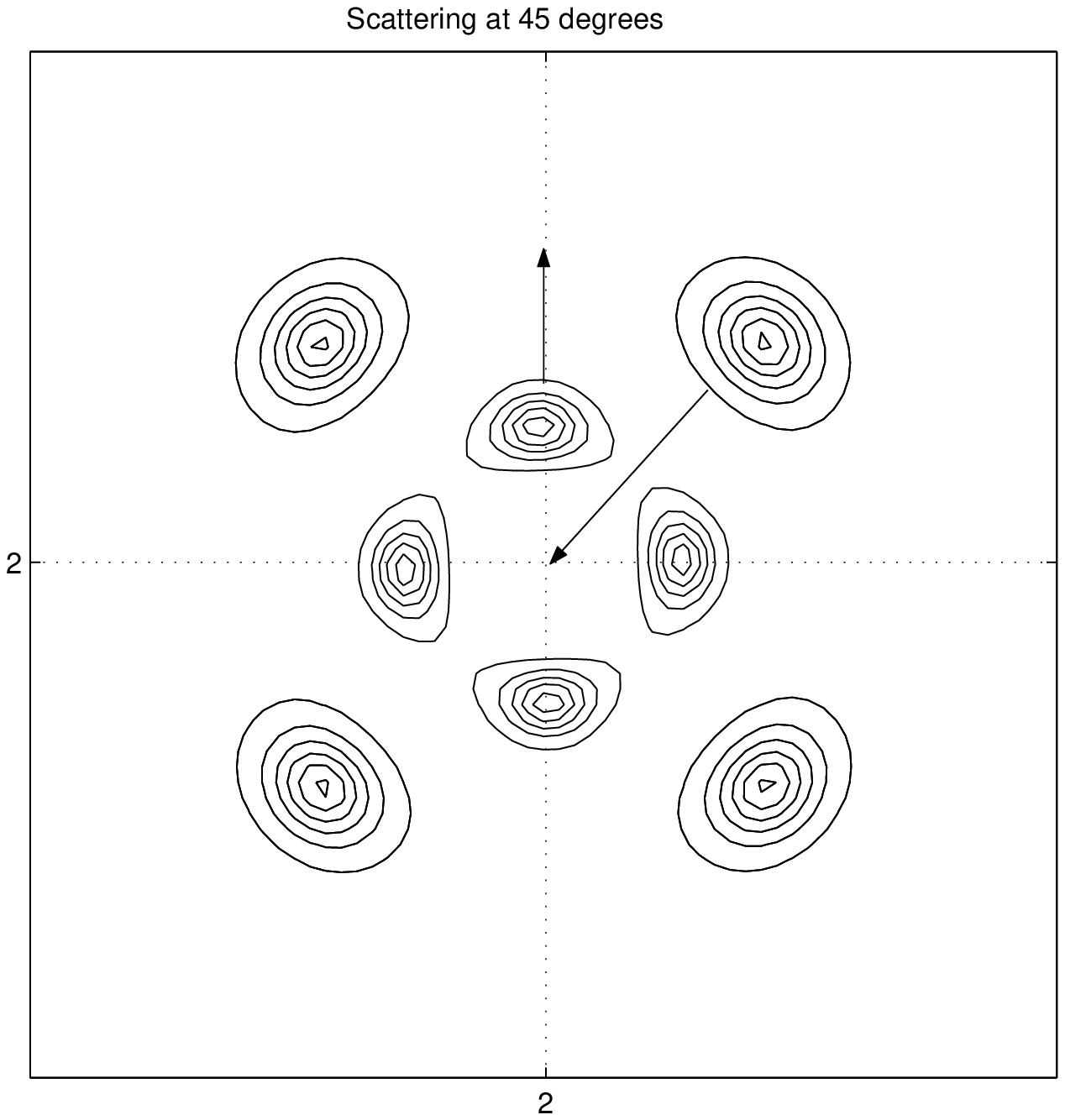}}
\caption{Scattering at $\pi/4$ radians.}
\label{fig:pic6n}
\end{figure}

The system (\ref{eq:4instanton}) gets moving via 
(\ref{eq:4boost}) and zeroes in on the middle of
the mesh, where the four skyrmions bump head-on into each other. This
dynamics makes the solitons scatter off and emerge towards the vertices of the 
dual square, emerging at $45^{\circ}$ with respect to 
the initial direction of motion, as depicted in diagram \ref{fig:pic5n}.  Figure
\ref{fig:pic6n} superimposes both the initial state ($t$=0) and the final state
($t=3$), allowing greater clarity in the appreciation of the scattering angle 
$\pi/4$.

Unlike the outcome of the previous section,
the dual square scattering on $T_2$ is not so surprising. For although 
the initial data doesn't have $D_4$ symmetry, it does have 
4-fold rotational symmetry and zero angular momentum. Note that
the boundary conditions (\ref{eq:boundaryw}) break the $SO(2)$ rotational symmetry
of the plane into a 4-fold rotational symmetry. 

\section{Geodesic approximation \label{sec:cc}}
Note that, in analogy to the $S_2$ case, when $v$ in (17) and (22) is set
to zero our initial configuration is a static solution of the equations
of motion. If we change $a_j$, $b_j$ to a new value given by (17) or (22)
for a particular value of $t$ the new configuration is again a 
static solution of the equations of motion. Hence as $v$ changes the changes 
(17) and (22) connect configurations which correspond to static solutions
of the equations of motion. Thus it is reasonable to expect that a system 
set off with a small $v$ will follow such a change. This expectation
goes under the name of geodesic approximation (the system evolves by changing
its zero-mode parameters). Its validity is not expected to depend too much
on whether the model is defined on $S_2$ or $T_2$.

So far we have been concerned with solitons of equal size. Let
us now look into the more general and interesting situation
of energy lumps of different sizes, illustrating 
the proceedings by studying the case $N=3$.

First we set up the initial configuration and 
evolve it through our standard numerical simulation. Then we choose
a set of collective coordinates to reproduce the results of our
simulations (trajectory, scattering), offering an explanation in
the framework of the geodesic approximation.

%
\subsection{Numerical simulation}    

Our 3-soliton system is still given by a
function of the form (\ref{eq:3instanton}), but 
with a layout not as symmetrical as before. Instead of 
(\ref{eq:rotation}) we put
%
\begin{equation}
\left.
\begin{array}{lll}
a'_j &=& a'_1 \exp{(i \alpha_j)} \\
b'_j &=& b'_1 \exp{(i \alpha_j)}
\end{array}
\right\}; \quad 
\alpha_j=(j-1)\frac{2\pi}{3}+(-1)^{(j-1)}(1-\delta_{1j})\xi,
\label{eq:rotationalpha}
\end{equation}   
with $j$=1,2,3 and the numbers $a'_1, \; b'_1$ being fixed
as customary. 
The initial three soliton configuration can be written as
%
\begin{eqnarray}
W &=& \frac{\sigma(z-a_1)}{\sigma(z-b)}
      \frac{\sigma(z-a_2)}{\sigma(z-b_2)}
      \frac{\sigma(z-a_3)}{\sigma(z-b_3)} \nonumber \\
  &=& \frac{\sigma(z'-a'_1)}{\sigma(z'-b')}
      \frac{\sigma(z'-a'_1 e\sp{i\alpha_2})} 
           {\sigma(z'-b'_1 e\sp{i\alpha_2})} 
      \frac{\sigma(z'-a'_1 e\sp{-i\alpha_2})}
           {\sigma(z'-b'_1 e\sp{-i\alpha_2})},
\label{eq:3instantonwithb}
\end{eqnarray}
where
%
\begin{equation}
b'=a'_1+2(a'_1-b'_1)\cos\alpha_2 \quad 
[ \alpha_2=\frac{2\pi}{3}-\xi ]
\label{eq:b}
\end{equation}
ensures that the zeros and poles in (\ref{eq:3instantonwithb}) comply
with the constraint in (\ref{eq:instanton}). As usual we switch between primed and
unprimed numbers through formula (\ref{eq:primedj}), plus $b'=b-c$ and
$z'=z-c$. We have also used the fact that
$\exp(i\alpha_3)=\exp(-i\alpha_2)$.

Note that equating $\xi$ to zero simplifies $\alpha_j$
to $(j-1) 2\pi/3 = \beta_j=$ of 
(\ref{eq:rotation}), whereupon the elliptic function
(\ref{eq:3instantonwithb}) reduces to the field
(\ref{eq:3instanton}) simply because \( \cos\alpha_2=\cos\beta_2=-1/2
\rightarrow b'=b'_1 \) according to (\ref{eq:b}).

An angle $\xi \ne 0$ generates solitons of different sizes which no
longer enjoy the positional symmetry boasted by the arrangement
(\ref{eq:rotation}). In the set-up (\ref{eq:rotation}), where
$\xi$=0 and the solitons have the same size, the required
relationship \( \sum a'_j = \sum b'_j \) looks after itself as 
shown in (\ref{eq:vaca}). For $\xi \ne 0$ a demand of the kind
(\ref{eq:b}) is needed.

Now we have evolved the 
solitons (\ref{eq:3instantonwithb}) 
in the pure version of the $CP^1$ model, with the solitons
being sent into collision in the regular fashion:
%
\begin{equation}
\left.
\begin{array}{lll}
a_j & \rightarrow & (a'_1 - v_0 t) \exp{(i\alpha_j)} + c \\
b   & \rightarrow & (b' - v_0 t) + c \\
b_s & \rightarrow & (b'_1 - v_0 t) \exp{(i\alpha_s)} + c \\
\end{array}
\right\};
\quad j=1,2,3 ; \; \; s=2,3.
\label{eq:boostesp}
\end{equation}  
 
The associated energy distribution is shown for a typical case by the contour 
plot of figure \ref{fig:pic7n}, which corresponds to a choice 
of parameters (\ref{eq:3values}) in addition to $\xi=10^{\circ}$.

The starting configuration is represented by the three wider structures
whereas the final, scattered state is the narrower trio plotted therein.
The sense of motion is clearly indicated by the arrows. We note that for
both configurations the solitons are rotated an extra angle $\xi$ with
respect to $W_1$, as compared to their partners of figure \ref{fig:pic3n},
And, unlike the latter, it is not the case that the three initial lumps
have the same size; nor they are situated at the vertices of an equilateral
triangle. Apparent as well from our simulations is 
that the scattering angle differs from $\pi/3$.  This is a consequence of 
considering nonsymmetrical solitons,
whose collisions are not elastic and thus involve energy transfer.

Clearly the distance between the solitons differ for different pairs of them
and so their interactions are not the same resulting in a different
scattering pattern. Can we explain this difference? In the next subsection
we show that an explanation can be provided in terms of,
appropriately chosen, collective coordinates and the motion 
following appropriate geodesics.

\subsection{Collective coordinates}

In order to make a wise selection of collective coordinates 
let us consider closely the positions and sizes of the solitons
defined in the previos subsection.

%
The location of the solitons (\ref{eq:3instantonwithb}) are 
clearly given by
\begin{equation}
{a'_1+b'\over 2}, 
\quad {a'_1+b'_1\over 2} e\sp{i\alpha_2},
\quad {a'_1+b'_1\over 2} e\sp{-i\alpha_2},
\label{eq:positions}
\end{equation}
and their sizes are
\begin{equation}
\vert{a'_1-b'\over 2}\vert,
\quad \vert{a'_1-b'_1\over 2}\vert,
\quad \vert{a'_1-b'_1\over2}\vert.
\label{eq:sizes}
\end{equation}

Put 
%
\begin{equation}
a'_1=k-\chi, \; b'_1=k+\chi 
\label{eq:chi}
\end{equation}
so the positions
(\ref{eq:positions}) adopt the form
%
\begin{equation}
k-\chi[1+2\cos(\alpha_2)], 
\quad k e\sp{i\alpha_2},
\quad k e\sp{-i\alpha_2},
\label{eq:positionschi}
\end{equation}
while the sizes (\ref{eq:sizes}) read
\begin{equation}
\vert 2 \chi \cos(\alpha_2) \vert,
\quad \vert \chi \vert, 
\quad \vert \chi \vert.
\label{eq:sizeschi}
\end{equation}
Consistently, for $\xi=0$ the description
(\ref{eq:positionschi})-(\ref{eq:sizeschi}) corresponds 
to objects of the same size $\vert \chi \vert$ which are
symmetrically situated at $k$ and $k \exp({\pm i 2\pi/3})$
(with respect to the centre). For simplicity, we have 
taken two lumps with the same size and a third lump with a different 
size. 

A possible collective coordinate description involves treating $k$, $\chi$ and
$\alpha_2$ as collective coordinates. Thus, in the simulation we expect $\alpha_2$
and $\chi$ to remain approximately constant and $k$ to vary. We are suggesting that
the scattering can be understood as proceeding (on average) with only $k$ depending
on time and varying from (take $k$ real for simplicity) $k>0$ to $k<0$. It is easy
to show, although a bit tedious in practice (this involves estimating various
elliptic integrals which can be done partially analytically, partially
numerically), that both on $\Re_2$ or on our torus the kinetic energy of the motion
just described is finite. So such a behaviour is possible.

We have carried out our collective coordinates ``motion'' using the 
specific values
\begin{equation}
\chi=(0.55, 0), \quad k=(u, 0), \; u \in [1,-1],
\label{eq:chivalue}
\end{equation}
where $u$ varies across the interval in steps of 0.2.  We see from
(\ref{eq:positionschi}) and (\ref{eq:chivalue}) that the configurations for
$u>0$ ($u<0$) correspond to incoming (outgoing/scattered) lumps. The value
$u \approx 0$ represents the situation where the solitons have collided and
are on top of each other, coalescing in the centre of the mesh.  It
is largely the behaviour of $W$ at $u \approx 0$ what determines the
scattering angle. 

\begin{figure}
\mbox{\epsfig{file=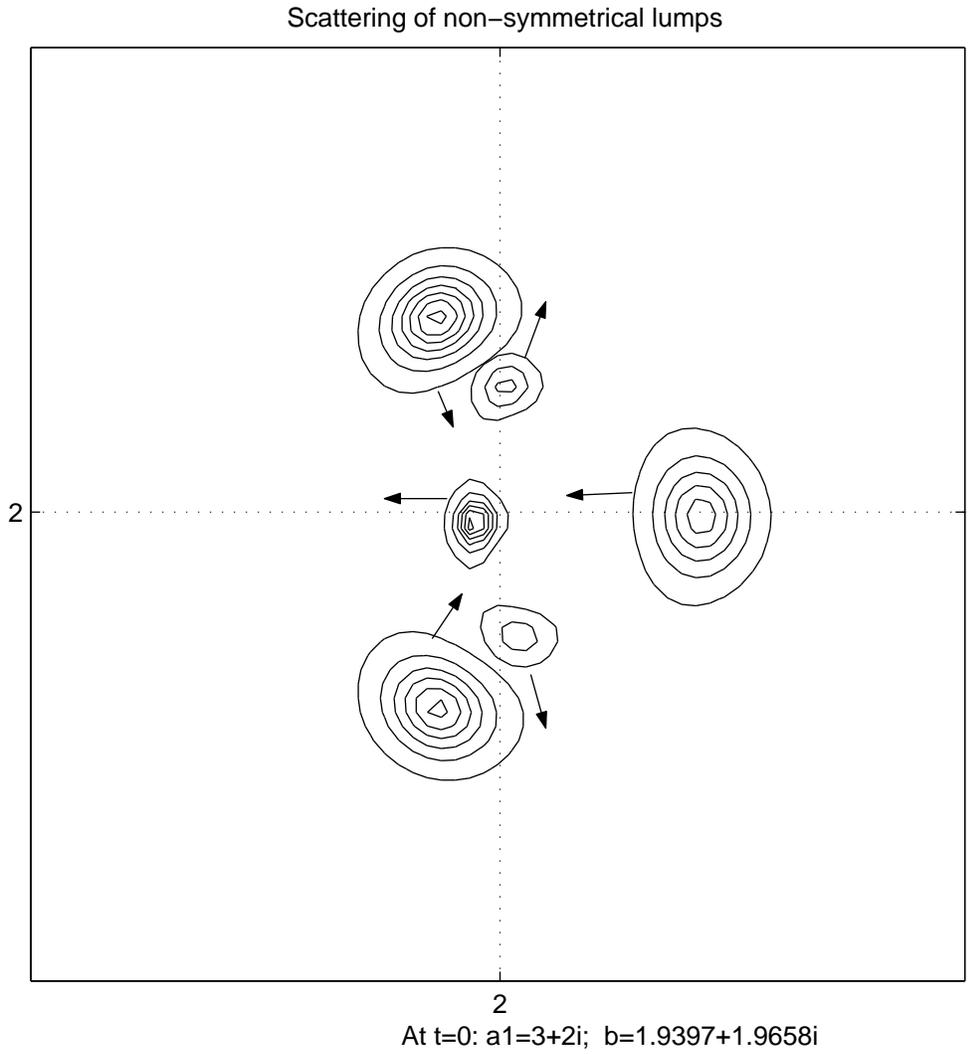}}   
\caption{Numerical simulation for solitons rotated 
an angle $\xi=10^{\circ}$
with respect to the symmetrical configuration of 
figure \ref{fig:pic1n}. They do not scatter at 
$60^{\circ}$.}
\label{fig:pic7n}
\end{figure}

\begin{figure}
\mbox{\epsfig{file=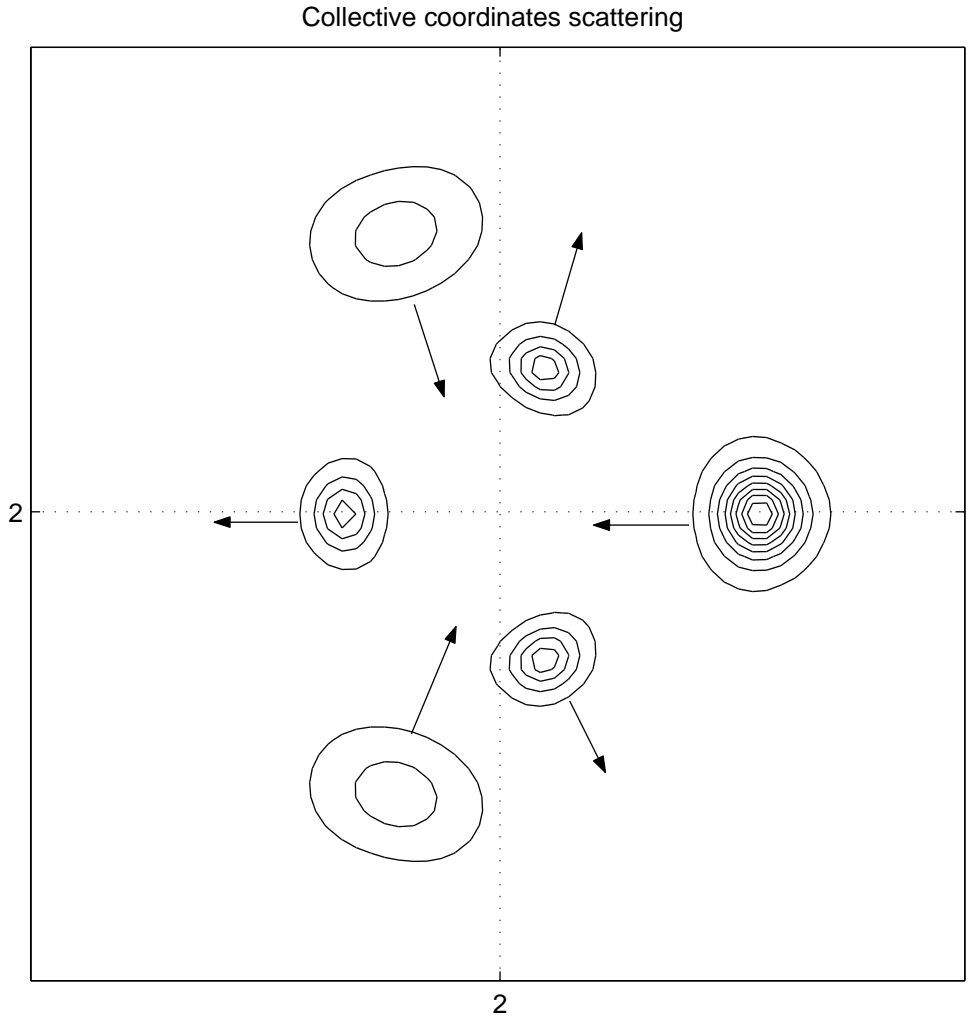}}
\caption{``Scattering'' according to our collective coordinates
view for $\xi=10^{\circ}$, to be compared with the numerical result shown in
figure \ref{fig:pic7n}. The incoming lumps are pictured for
$k=(1,0)$ and the outgoing structures correspond to $k=(-0.2,0)$. It is not
important that the widths of these lumps are not the same as their sibblings
of figure \ref{fig:pic7n}: we worry about the relative positions and scattering
angle.}
\label{fig:pic8n}
\end{figure}

The state for $k$=(1,0) (incoming lumps) and $k$=(-0.2,0) (``scattered'' lumps) are
depicted in figure \ref{fig:pic8n} for the case $\xi=10^{\circ}$. As long as the
path followed by the lumps and the scattering angle is concerned, the similarity
with the numerically evolved situation of figure \ref{fig:pic7n} is clear. We
stress that since we are mostly interested in the relative position of the solitons
and their scattering angles, it is immaterial that the breadths of the solitons in
figure \ref{fig:pic7n} differ from the solitons of figure \ref{fig:pic8n}.

A more detailed comparison can be made by viewing a trajectory plot, one showing
consecutive snapshots [corresponding to $k$=(1,0),
(0.8,0),...,(-0,2),(-0.4,0)...(-1,0)]
of lump positions as they approach each other and scatter off.  

\begin{figure}
\mbox{\epsfig{file=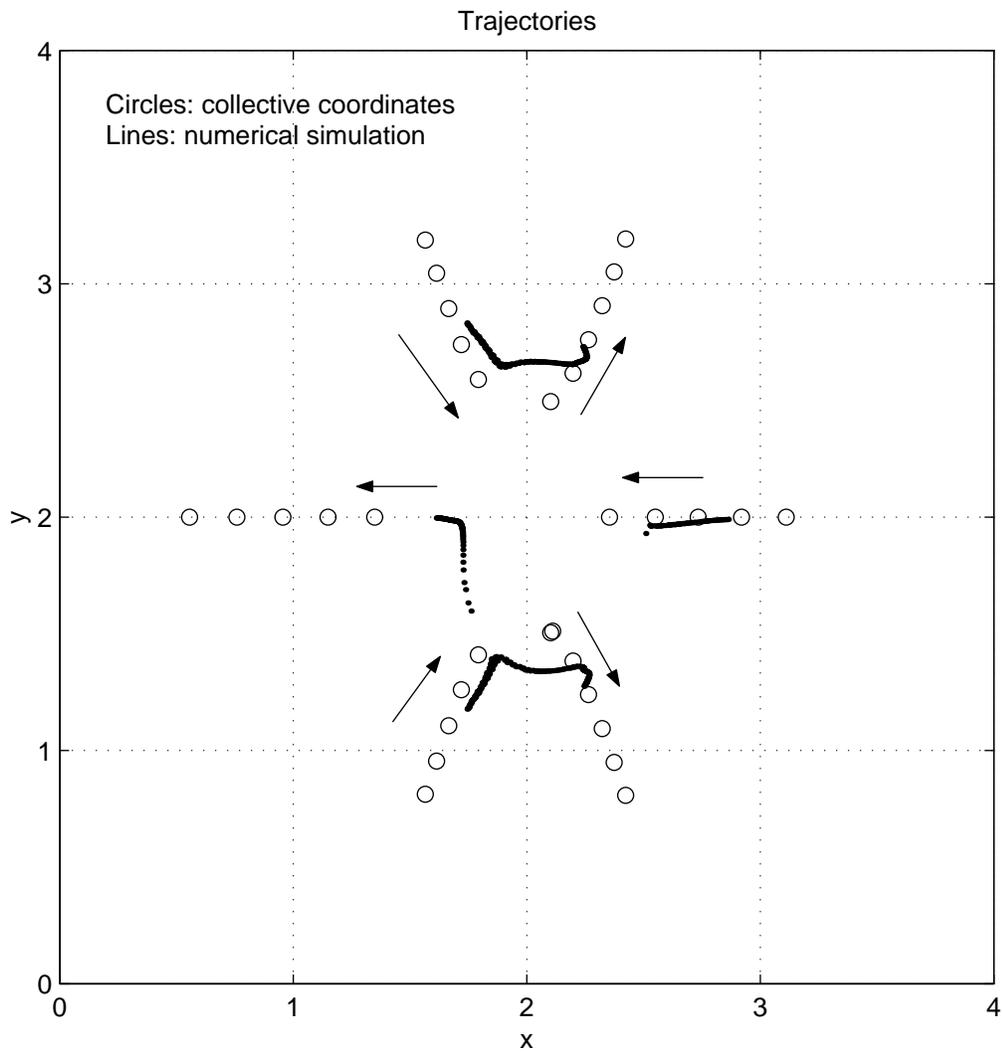}}
\caption{Trajectory plots ($\xi=10^{\circ}$) sketching the path 
followed by the
humps presented in figures \ref{fig:pic7n} and \ref{fig:pic8n}.
The itinerary according to our geodesic approximation (circles)
shows very good agreement with the motion obtained via
numerical simulation (solid lines).}
\label{fig:pic9n}
\end{figure}

This is shown in figure \ref{fig:pic9n}, where the full collective coordinate motion
has been ticked with small circles and the motion according to the time evolution of
subsection 6.1 has been sketched with solid lines.  Note that after scattering the
path of the numerically evolved lumps (continuos lines) cannot be followed much
farther; this is because the solitons get very spiky and the simulation breaks down.

\begin{figure}
\mbox{\epsfig{file=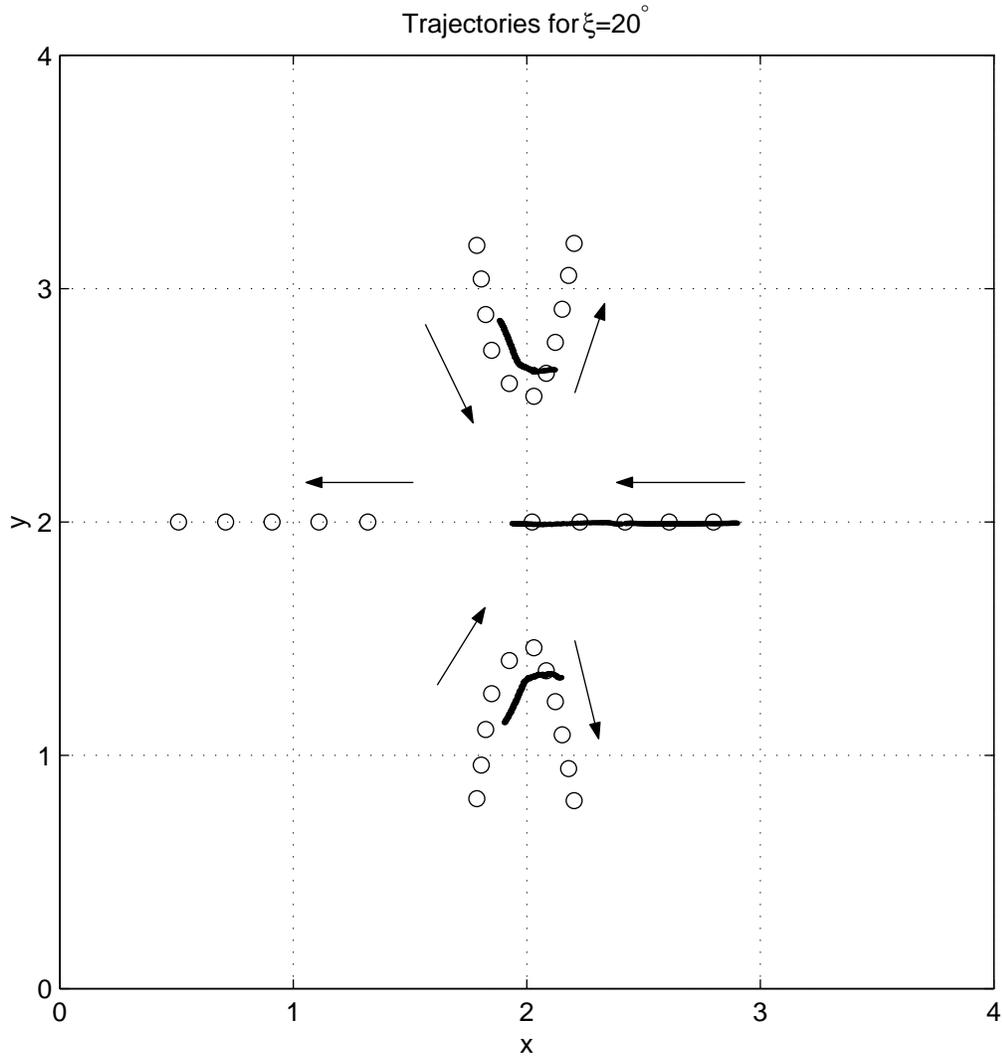}}
\caption{Plots comparing the trajectory obtained via 
numerical simulations (solid lines) and the trajectory 
gotten using the geodesic approximation (circles) for
$\xi=20^{\circ}$.}
\label{fig:pic10n}
\end{figure}

Our simulations have been carried out for several values of $\xi$ and have been
compared with the corresponding collective coordinate motion. We have always found
the trajectories from both approaches to be in agreement, thus supporting our choice
of collective coordinates. Another illustration is provided by figure
\ref{fig:pic10n} for the case $\xi=20^{\circ}$.

\newpage
\section{Conclusions \label{sec:conclusions}}

In this article we have studied head-on collisions between solitons
in the (2+1)-D $CP^1$ model with periodic boundary conditions, that
is, with the model defined on a flat torus $T_2$. 

Through numerical simulations we have found that solitons of equal sizes
(at the initial time) scatter at an angle $\pi/N$, or dual-polygon scattering,
where $N$ is the soliton  number or topological charge of the system. In this paper
we have focused on $N=3,4$ (the case $N=2$ has been considered previously \cite{nonl}
and it has been found to comply with the $\pi/2$ scattering). 

Unlike the usual model on $\Re_2$, 
our model with periodic boundary conditions breaks the $SO(2)$ rotational
invariance of the plane, leaving us with a numerical mesh with directed sides where
the initial soliton configuration has no symmetry under the dihedral group $D_N$.
So it is remarkable to still find dual-triangle scattering in this format. On the
other hand, since the $SO(2)$ symmetry breaks into a four-fold rotational symmetry,
the dual-square scattering is less unexpected.

We have also considered solitons of different sizes 
(at the initial time) and have observed that the scattering angle
in no longer $\pi/N$, outcome arising from the fact that there is 
energy transfer in collisions between unsymmetrical solitons. 

By reparametrising the quantities describing the positions of the 
solitons using a juditious set of collective coordinates, we have been
able to reproduce the above numerical results, thus offering an explanation
of the scattering process. We have illustrated this approach using a
3-soliton field.   

These results raise important questions pertaining to the               
the interplay between the symmetry of the initial
configuration and the lack of symmetry of the torus itself. As pointed out
at the end of section \ref{sec:3}, the non-isotropy of the torus might affect
the evolution of the lumps if they are initially placed 
near the boundary of the mesh. What about systems with $N$=5,6,...? Numerical 
experiments on a periodic, rectangular grid would also be worth performing.
We hope to report on these matters in the near future. 
 
\vspace{10 mm}
\Large{\bf Acknowledgements} \\

\normalsize
Work for this paper was carried out during the 
authors' visits to Durham and Maracaibo, supported by the ROYAL 
SOCIETY, CONICIT and LUZ-CONDES. The paper was completed 
at IISC, Bangalore-India, where $\Re$J Cova stayed thanks 
to a grant TWAS-UNESCO. He thanks Prof Uberoi and 
Prof Rangarajan for their hospitality at IISc.



\begin{thebibliography}{99}
%
\bibitem{cho} Proc. \emph{Physics in (2+1) dimensions (Korea),
World Scientific} (1992)
%
\bibitem{leese} Leese RA \emph{et al} \emph{Nonlinearity} \textbf{3},
387 (1990)
\bibitem{skyrme} Skyrme THR \emph{Nucl Phys} {\bf 31}, 556 (1962)
%
%
\bibitem{nonl} Cova RJ and Zakrzewski WJ {\em Nonlinearity} {\bf 10},
1305 (1997)
%
%
\bibitem{martin} Speight JM \emph{Comm Math Phys} \textbf{194}, 513
(1998)
%
\bibitem{epj2k1} Cova RJ \emph{Eur Phys Jour B} \textbf{23}, 201
(2001)
%
\bibitem{pin} Kudryavtsev A, Piette B and Zakrzewski W
\emph{Phys Lett A} \textbf{183}, 119 (1993)
%
%
\bibitem{goursat} Goursat E \emph{Functions of a complex variable},
Dover Publications (1916)
%
\bibitem{lawden} Lawden DF {\em Elliptic functions and applications}
Springer Verlag (1989)            
%
\bibitem{mcgraw} Erd\'{e}lyi A {\em et al} 
{\em Higher transcendental functions} vol II Mc Graw Hill (1953).
%
\bibitem{chaos} Piette B and Zakrzewski W \emph{Chaos, solitons and
fractals} {\bf 5}, 2495 (1995)

\end{thebibliography}
\end{document}